\documentclass[twocolumn,secnumarabic,amssymb, nobibnotes, aps, prd]{revtex4}
\usepackage{graphicx}

\begin{document}
\title{Comment on "Case of thermodynamic failure in the Ginzburg-Landau approach to fluctuation superconductivity"}
\author{A.V. Nikulov}
\email[]{nikulov@iptm.ru}
\affiliation{Institute of Microelectronics Technology and High Purity Materials, Russian Academy of Sciences, 142432 Chernogolovka, Moscow District, RUSSIA.} %nikulov@ipmt-hpm.ac.ru
%\date{}
\begin{abstract} Jorge Berger shows theoretically in the paper Phys. Rev. B 109, 024501 (2024) that according to the Ginzburg-Landau theory the persistent current can create the persistent voltage, i.e. a dc voltage at thermodynamic equilibrium, on segments of nonuniform superconducting loop. A similar result was published early and was collaborated experimentally. The persistent power estimated by Berger is compared with the experimentally observed power. The attention of experimenters is drawn to the possibility to observe the persistent voltage thanks to its increase with the number of identical rings connected in series.
\end{abstract}

\maketitle 

\narrowtext

\section{Introduction}
Jorge Berger has shown in \cite{Berger2024PRB} that the persistent current of Cooper pairs $I_{p}$ can create a persistent voltage $V_{p}$, i.e. the dc voltage $V_{dc}$ at thermodynamic equilibrium, on segments of nonuniform superconducting loop. This result deserves special attention because of its contradiction with the second law of thermodynamics. It is written in the review article \cite{PhysRep1999}: "{\it The second law of thermodynamics is, without a doubt, one of the most perfect laws in physics. Any reproducible violation of it, however small, would bring the discoverer great riches as well as a trip to Stockholm. The world's energy problems would be solved at one stroke. It is not possible to find any other law (except, perhaps, for super selection rules such as charge conservation) for which a proposed violation would bring more skepticism than this one. Not even Maxwell's laws of electricity or Newton's law of gravitation are so sacrosanct}". 

It is important to draw reader's attention to the publications which confirm experimentally and theoretically the result published in \cite{Berger2024PRB}, because of this skepticism about violation of the sacrosanct law. Its reproducible violation can solve world's energy problems since we must used a fuel in our heat engine because of the Carnot principle which we call "{\it as the second law of thermodynamics since Clausius's time}" \cite{Smoluchowski}. The efficiency of any heat engine cannot exceed the maximum value 
$$\eta_{max} = 1 - \frac{T_{co}}{T_{he}}    \eqno{(1)}$$
which is determined by the ratio $T_{co}/T_{he}$ of the cooler temperature $T_{co}$ to the heater temperature $T_{he}$ according to the Carnot principle. Therefore we must create and maintain a temperature difference $T_{he} - T_{co}$ with the help of a fuel in order to obtained a useful work from heat.

%We must used a fuel in our heat engine in order to create and to maintain a temperature difference $T_{he} - T_{co}$ since the efficiency $\eta = W/\Delta Q$ of any heat engine is zero $\eta  = 0$ without a temperature difference $T_{he} - T_{co} = 0$, according to the Carnot principle (1), which we call as the second law of thermodynamics since Clausius's time. A violation of the second law of thermodynamics can solve the world's energy problems since we will be able in this case convert the heat energy $\Delta Q$ to an useful work $W$ without any fuel. The theoretical result obtained by Berger \cite{Berger2024PRB} suggests such a possibility. Therefore it is very important to evaluate this possibility and the reliability of the result obtained in \cite{Berger2024PRB} by comparing it with other published results, primarily experimental ones.

\section{A useful work from heat without any fuel}
The theoretical result \cite{Berger2024PRB} allows to convert the heat energy $\Delta Q$ to a useful work $W$ without any fuel. According to Berger's calculations the potential electric field $E(x) = -dV/dx$ at thermodynamic equilibrium creates the persistent voltage on segments $A$ $\int_{A}dx E(x) = -V_{p}$ and $B$ $\int_{B}dx E(x) = V_{p}$ of the loop $L = A + B$. The dc voltage are equal in value and opposite in sign since $\int_{L}dx E(x) = \int_{L}dx (-dV/dx) \equiv 0$ \cite{Berger2024PRB}. The sign of the persistent voltage $V_{p}$ and the persistent power $V_{p}I_{p}$ is determined relative to the direction of the persistent current $I_{p}$ circulating in the loop $L = A + B$. The segment $A$ in which $V_{p}I_{p} < 0$ is the power source and the segment $B$ in which $V_{p}I_{p} > 0$ is the load. 

The persistent voltage $V_{p}$ can create a current $I_{load}$ in a useful load connected to the segment $A$. The total power $-VI_{A} + VI_{B} < 0$ delivered by the electric field in the whole loop will be negative in this case since the $I_{A}$ current in the $A$ segment should be equal the sum $I_{A} = I_{B} + I_{load}$ and $\int_{L}dx E(x) = \int_{L}dx (-dV/dx) \equiv 0$ in any case. A useful work $W = \int_{t}dt V_{p}I_{load}$ can be obtained in the load from heat without the need to create and to maintain a temperature difference $T_{he} - T_{co}$. Thus, the question of the possibility of the persistent voltage $V_{p}$ is of great fundamental and practical importance.   

\section{A possibility of the persistent voltage}
This possibility may be expected from experimental evidence of the persistent current $I_{p}$ observed at a non-zero resistance $R_{n} > R > 0$ in the fluctuation region near superconducting transition $T \approx T_{c}$ \cite{LP1962,Science2007,Letter2007,toKulik2010}. It is known that the conventional electric current $I$ creates the potential difference
$$V = 0.5(R _{B} - R _{A})I   \eqno{(2)}$$                                                        
on the halves of the ring with different resistance $R _{B} > R _{A}$. The value and the direction (clockwise or counterclockwise) of the persistent current change periodically with magnetic field $I_{p}(\Phi /\Phi _{0})$ with the period $B_{0} = \Phi _{0}/\pi r^{2}$ corresponding to the flux quantum $\Phi _{0} = 2\pi \hbar /2e \approx 20.7 \ Oe \ \mu m^{2}$ inside the ring with the radius $r$, according to experimental results \cite{Science2007,Letter2007,toKulik2010}. 

Observations \cite{Letter2007,toKulik2010,PM2001,Letter2003,PLA2012PV,APL2016,PLA2017} on halves of asymmetric superconducting rings of the quantum oscillations of the dc voltage $V_{dc}(\Phi /\Phi _{0}) \propto I_{p}(\Phi /\Phi _{0})$ give experimental evidence that persistent current can also create the potential difference. The persistent current flows against the total electric field $E = -\nabla V_{p}$ in the $A$ segment according to these observations since the Faraday electric field $dA/dt = 0$ at a magnetic flux inside the ring $\Phi = \pi r^{2}B \neq n\Phi _{0}$ constant in time $d\Phi /dt = 2\pi r dA/dt = 0$ \cite{Physica2019}. 

\section{The Ginzburg-Landau theory explains paradoxical phenomena}
The quantum oscillations of the dc voltage $V_{dc}(\Phi /\Phi _{0}) \propto I_{p}(\Phi /\Phi _{0})$ were observed first in 1967 at the measurements of an asymmetric dc SQUID \cite{Physica1967}. The dc power $P_{dc} = V_{dc}^{2}/R$ observed in \cite{Physica1967} and also in \cite{Letter2007,toKulik2010,PM2001,Letter2003,PLA2012PV,APL2016,PLA2017,Physica2019} can be used for a useful work $W = \int_{t}dt P_{dc}$. The authors \cite{Physica1967} concluded that the energy for this work is taken from a non-equilibrium noise since they stopped observing the dc voltage $V_{dc}(\Phi /\Phi _{0}) \propto I_{p}(\Phi /\Phi _{0})$ after careful shielding of all parts of their equipment. They did not taken into account that the persistent current $I_{p}(\Phi /\Phi _{0}) \propto V_{dc}(\Phi /\Phi _{0})$ flowing against the total dc electric field $E = -\nabla V_{dc}$ is a paradoxical phenomenon and that the persistent current observed at a non-zero resistance $0 < \overline{R} < R_{n}$ in the Little-Parks experiment \cite{LP1962} implies the need for a DC power source $\overline{RI_{p}^{2}}$. 

The GL theory \cite{GL1950} explains the both paradoxical phenomena as a consequence of the quantization of the canonical momentum of $\oint_{l}dl p = \oint_{l}dl (mv + qA) = n2\pi \hbar $ and with the help of postulate that all $N_{s} = n_{s}V = n_{s}s2\pi r$ Cooper pairs have the same quantum number $n$. The average persistent current $\overline{I_{p}} = \int_{\Theta }dtI(t)/\Theta  = q\overline{N_{s}}\overline{v}/2\pi r$ does not damped at $0 < \overline{R} < R_{n}$ since the velocity of Cooper pairs should increase from $v  =  0$ to 
$$\oint_{l}dl v  =  \frac{2\pi \hbar }{m}(n - \frac{\Phi}{\Phi_{0}}) \eqno{(3)}$$
each time when the ring returns to the superconducting state at its switching by thermal fluctuations with a frequency $f_{sw} \gg 1/\Theta $. The total change of the momentum $mv$ because of the quantization (3) during a time unity $F_{q} = \hbar (\overline{n} - \Phi /\Phi_{0})f_{sw}/r$ was called 'quantum force' in \cite{PRB2001}. Thus, according to the GL theory the persistent current $\overline{I_{p}} \neq 0$ is not damped at $0 < \overline{R} < R_{n}$ \cite{LP1962,Science2007,Letter2007,toKulik2010} due to the force balance $\overline{F_{dis}} + F_{q} = 0$: the average velocity of electrons should decrease from $v = (\hbar /mr)(n - \Phi /\Phi_{0})$ to $v = 0$ under influence of the dissipation force $F_{dis} = -\eta \overline{v}$ because of electron scattering, but the velocity of Cooper pairs should increase from $v  =  0$ to $v  =  (\hbar /mr)(n - \Phi /\Phi_{0})$ because of the quantization (3).   

A result similar to the one obtained in \cite{Berger2024PRB} was obtained in the article \cite{LTP1998} for a ring whose halves have different critical temperatures $T_{c,B} < T_{c,A}$. The GL theory predicts the persistent voltage at $T \approx T_{c,B} < T_{c,A}$ since only the $B$ half  should be switched by thermal fluctuations \cite{LTP1998}. The persistent voltage $V _{p} = 0.5(R _{B} - R _{A})\overline{I_{p}} = 0.5R _{B}\overline{I_{p}}$ may be expected also according observations of the persistent current both at $T < T_{c}$ where $R = 0$ \cite{JETP07} and at $T \approx T_{c}$ where $0 < R < R_{n}$ \cite{LP1962,Science2007,Letter2007,toKulik2010}. 

The dc voltage $V_{dc} \approx  Lf_{sw}\overline{I_{p}}$ should be observed when the $B$ half is switched with a low frequency $f_{sw} \ll 1/\tau_{RL}$ and $V_{dc} \approx  R_{B}\overline{I_{p}}$ with a high frequency $f_{sw} \gg  1/\tau_{RL}$ \cite{LTP1998,PRA2012QF} since the potential difference $V(t) = R_ {n,B}I(t) = R_{n,B}I_{p}\exp -t/\tau_{RL}$ should appear on the $B$ half after each transition to the normal state with the resistance $R_ {n,B}$, see Fig.2 in \cite{Physica2021}. $\tau_{RL} = L/R_ {B}$ is the relaxation time. It should be noted that according to the GL theory, the amplitude of $V_{dc}(\Phi /\Phi _{0})$ depends on the amplitude of $I_{p}(\Phi /\Phi _{0})$, and not on the amplitude of the non-equilibrium noise, as the authors \cite{Physica1967} thought. 

The experimental observations $\overline{I_{p}}$ in the fluctuation region \cite{LP1962,Science2007,Letter2007,toKulik2010} were explained in \cite{PRB2001} as a directed (non-chaotic) Brownian motion. M. Smoluchowsk realized a hundred and ten years ago that the impossibility of a directed Brownian motion is one of the conditions for the validity of the second law of thermodynamics \cite{Smoluchowski}. He was right! Such well known chaotic Brownian motion \cite{Feynman} as the Nyquist \cite{Nyquist} (or Johnson \cite{Johnson}) noise cannot be used for a useful work since its power $\overline{V_{Nyq}^{2}}/R = 4k_{B}T\Delta f $ \cite{Feynman} is the same for all elements of electrical circuit in any frequency band $\Delta f $. In contrast to this we obtain the power source $V_{p}I_{p} < 0$ and the load $V_{p}I_{p} > 0$ according to \cite{Berger2024PRB,LTP1998} in the case of such directed Brownian motion as the persistent current \cite{LP1962,Science2007,Letter2007,toKulik2010}.

Berger writes \cite{Berger2024PRB} that he considers a situation beyond the realms considered in the articles \cite{Hirsch2020EPL} and \cite{Physica2021}. It is not quite right. The well-known physicist W.H. Keesom wrote in 1934 that "{\it it is essential that the persistent currents have been annihilated before the material gets resistance, so that no Joule-heat is developed}" \cite{Keesom1934} since he understood that the emergence of the persistent currents at the Meissner effect \cite{Meissner1933} contradicts to the second law of thermodynamics in the opposite case. The macroscopic persistent current $I_{p} = qN_{s}v/2\pi r$ with the macroscopic kinetic energy $E_{p} = N_{s}mv^{2}/2 \gg k_{B}T$ emerges also at each transition to the superconducting state of a ring, for example during measurements of the critical current \cite{PRB2014}. Jorge Hirsch draws attention in \cite{Hirsch2020EPL} that the conventional theory of superconductivity \cite{BCS1957} predicts that Joule heat is generated when the persistent currents decrease. He states that this theory should be questioned because of its contradiction with the laws of thermodynamics. But rather the laws of thermodynamics than the theory \cite{GL1950,BCS1957} should be questioned \cite{Physica2021}, because of numerous experimental evidences  \cite{LP1962,Science2007,Letter2007,toKulik2010} that the persistent currents are not annihilated not only before but even after the material gets resistance.  

\section{Observation of the persistent voltage}
The DC power is summed $(NV_{dc})^{2}/NR = NV_{dc}^{2}/R$, as opposed to the chaotic power of the Nyquist noise $\overline{NV_{Nyq}^{2}}/NR = \overline{V_{Nyq}^{2}}/R = 4k_{B}T\Delta f $. The measurements \cite{Letter2007,PM2001,Physica2019} have corroborated the $V_{dc}(\Phi /\Phi _{0})$ increase with the number $N$ of identical asymmetric ring connected in series $V_{A,N} = NV_{A,1}$ \cite{Physica2019}. This advantage allowed to conform \cite{Letter2007,toKulik2010,PM2001,PLA2012PV,APL2016} that the dc voltage $V_{dc}(\Phi /\Phi _{0})$ do not disappear after careful shielding of a non-equilibrium noise but their amplitude $V_{A}$ decreases below the detection limit $V_{A} < V_{A,dl} $ on single ring. In accordance with the GL theory the amplitude $V_{A}$ depends on the amplitude of the persistent current $I_{p,A}$ which decreases with the temperature increase  \cite{JETP07}
$$I_{p,A}(T) =  I_{p,A}(T = 0)(1 - T/T_{c})  \eqno{(4)}$$ 
The amplitude $V_{A}$ decreases after careful shielding \cite{Physica1967} since a non-equilibrium noise \cite{Letter2007,toKulik2010,PM2001} or an external ac current \cite{PLA2017,Physica2019,Letter2003,JETP07} cannot switch the ring in the normal state at $T < T_{c}$ when their amplitude $I_{A} = \overline{2I _{noise}^{2}}^{1/2}$ does not exceed the critical current 
$$I_{c}(T) =  I_{c}(T = 0)(1 - T/T_{c})^{3/2}  \eqno{(5)}$$ 

The persistent voltage $V_{p}(\Phi /\Phi _{0})$, i.e. the dc voltage at thermodynamic equilibrium, can be observed only in a narrow temperature region corresponding the resistive transition $R_{n} > R(T) > 0$ where thermal fluctuations decrease the critical current (5) down to zero $I_{c}(T) = 0$ at $T < T_{c}$ and makes the persistent current (4) non-zero $I_{p,A}(T) > 0$ at $T > T_{c}$ \cite{Skocpol1975}. No non-equilibrium noise is needed for ring switching in this region. A system of 1080 asymmetric aluminum rings connected in series was needed in order to observe the persistent voltage $V_{p}(\Phi /\Phi _{0})$ with the amplitude exceeding the detection limit $V_{A} < V_{A,dl} \approx 20 \ n V = 2 \ 10^{-8} \ V$ in the temperature region $1.359 \ K < T < 1.373 \ K $ corresponding to the lower part of the resistive transition $0.01R_{n} < R(T) < 0.32R_{n}$ \cite{PLA2012PV}. 

The observation of the persistent power in \cite{PLA2012PV} is a qualitative confirmation of the theoretical result \cite{Berger2024PRB}, despite the fact that the maximum power $P \approx 5 \ 10^{-12} \ W/m$ estimated on the base of experimental results \cite{PLA2012PV} is less than the power $P \approx 10^{-10} \ W/m$ theoretically estimated by Berger \cite{Berger2024PRB}. The maximum amplitude $V_{A,1080} \approx 0.14 \ \mu  V = 1.4 \ 10^{-7} \ V$ observed on the 1080 rings corresponds to $V_{A,1} \approx 1.4 \ 10^{-10} \ V$ on one ring if each ring with the length of semi-ring $L/2 = \pi r \approx 2.8 \ \mu m = 2.8 \ 10^{-6} \ m $ contributes to the observed power. The amplitude of the persistent $I_{p}(\Phi /\Phi _{0})$ equals approximately $I_{p} \approx 0.1 \ \mu A = 10^{-7} \ A$ at the temperature corresponding to the resistance $R(T) \approx  0.2R_{n}$ for the rings used in \cite{PLA2012PV} according to the measurements of the Little-Parks oscillations \cite{Letter2007}. The maximum persistent power $V_{p}I_{p} \approx V_{A,1}I_{p,A} \approx 1.4 \ 10^{-10} \ V \times 10^{-7} \ A \approx 1.4 \ 10^{-17} \ W$ per one ring observed in \cite{PLA2012PV} is much less than the dc power $P_{dc} \approx 2 \ 10^{-10} \ W$ observed in \cite{Physica1967} and $P_{dc} \approx 10^{-12} \ W$ observed in \cite{PM2001} since the persistent current $I_{p}(\Phi /\Phi _{0})$ in the fluctuation region is much smaller than in the superconducting state.    

Non-equilibrium noise was suppressed by additional filtration with the help of the help of low-temperature $\pi $-filters and coaxial resistive twisted pairs in order to observe the persistent voltage $V_{p}(\Phi /\Phi _{0})$ in \cite{PLA2012PV}. This additional filtration allowed also to demonstrate that the system of a big number of identical asymmetric superconducting rings can be used as a detector of a noise: the oscillations $V_{dc}(\Phi /\Phi _{0})$ with the amplitude up to $V_{A,667} \approx 1 \ \mu  V = 1 \ 10^{-6} \ V$ were observed on a system of 667 asymmetric aluminum rings with $r \approx 0.5 \ \mu m$ (the amplitude $V_{A,1} \approx 1.5 \ n  V = 1.5 \ 10^{-9} \ V$ on each ring) when a source of a noise with the amplitude $\overline{2I _{noise}^{2}}^{1/2} \approx 50 \ nA = 5 \ 10^{-8} \ A$ was switched on \cite{APL2016}. The non-equilibrium noise $\overline{2I _{noise}^{2}}^{1/2} \approx 5 \ 10^{-8} \ A$ is weaker than the noise  $\overline{2I _{noise}^{2}}^{1/2} \approx 20 \ \mu A = 2 \ 10^{-5} \ A$ induced $V_{dc} \approx 1.5 \ 10^{-5} \ V$ in \cite{Physica1967} and the noise $\overline{2I _{noise}^{2}}^{1/2} \approx 3 \ 10^{-6} \ A$ induced $V_{dc}(\Phi /\Phi _{0}) \approx 10^{-6} \ V $ in \cite{PM2001}. 

\section{Conclusion} 
The persistent power of one ring cannot exceed in any case the total power of the Nyquist \cite{Nyquist} (or Johnson \cite{Johnson}) noise $V_{p}^{2}/R < \overline{V_{Nyq}^{2}}/R  = 4k_{B}Tf_{ql} = 4(k_{B}T)^{2}/h$ in the whole frequency band from $f = 0$ to the quantum limit $f = f{ql} \approx k_{B}T/h$ \cite{Feynman}. More than $10^{13}$ of the rings with $T_{c} \approx 1 \ K$ used in \cite{PLA2012PV} are needed in order to get a power of only 1 W since $4(k_{B}T)^{2}/h \approx 10^{-13} \ W$ at $T \approx 1 \ K$. Fewer rings $> 10^{9}$ will be required if high-temperature superconductors (HTSC) with $T_{c} \approx 100 \ K$ can be used. Even the use of HTSC rings is unlikely to be able to solve the world's energy problems at one stroke. But the fact of violation of the second law of thermodynamics is important in itself, since if the violation is possible in one case, then it may be possible in other cases, some of which may be more perspective for practical application. 

One example of the violation may be enough to make scientists stop to consider the second law of thermodynamics as sacrosanct law. In this case, the world's energy problems may be solved if not at one stroke, but rather quickly. Therefore, it is important to draw the attention of experimenters to the possibility to observe the persistent voltage not only in the lower part of the resistive transition, but in the entire temperature region of thermal fluctuations with the help of a system with a number of rings significantly exceeding 1000.

This work was made in the framework of State Task No 075-00296-24-00.

\end{document}